\documentclass[a4paper, 10pt, conference]{ieeeconf}
\usepackage{algorithmic}
\usepackage{algorithm}
\usepackage{multirow}
\usepackage{tikz}
\usepackage{amsmath}
\usepackage{booktabs}
\usepackage{graphicx}
\usetikzlibrary{calc}
\usepackage{pgfplots}
\usetikzlibrary{calc}
\pgfplotsset{compat=1.15}

\AtBeginDocument{
  
}

\IEEEoverridecommandlockouts

\title{\LARGE \bf
Exploring the Temporal Dynamics of Facial Mimicry in Emotion Processing Using Action Units}

 \author{
     \parbox{16cm}{\centering
     {\large Meisam J. Seikavandi$^1$, Jostein Fimland$^1$, Maria Barrett$^2$, and Paolo Burelli$^1$}\\
     {\normalsize $^1$ brAIn lab, IT University of Copenhagen, Denmark \\
     $^2$ IT University of Copenhagen, Denmark}
     }
     }

\begin{document}
\maketitle
\thispagestyle{empty}
\pagestyle{empty}
\vspace{-10pt}

\begin{abstract}
Facial mimicry—the automatic, unconscious imitation of others' expressions—is vital for emotional understanding. This study investigates how mimicry differs across emotions using Face Action Units from videos and participants' responses. Dynamic Time Warping quantified the temporal alignment between participants' and stimuli's facial expressions, revealing significant emotional variations. Post-hoc tests indicated greater mimicry for 'Fear' than 'Happy' and reduced mimicry for 'Anger' compared to 'Fear'. The mimicry correlations with personality traits like Extraversion and Agreeableness were significant, showcasing subtle yet meaningful connections. These findings suggest specific emotions evoke stronger mimicry, with personality traits playing a secondary role in emotional alignment.
Notably, our results highlight how personality-linked mimicry mechanisms extend beyond interpersonal communication to affective computing applications, such as remote human-human interactions and human-virtual-agent scenarios. Insights from temporal facial mimicry—e.g., designing digital agents that adaptively mirror user expressions—enable developers to create empathetic, personalized systems, enhancing emotional resonance and user engagement.
\end{abstract}

\section{Introduction}  

Facial mimicry, the automatic imitation of others' expressions, plays a crucial role in empathy and social interaction~\cite{hatfield1994,dimberg2000}. Often termed the `social glue'~\cite{lakin2003chameleon}, it fosters emotional connection and social cohesion. This subtle, unconscious reaction is central to emotion processing~\cite{oberman2007face,ponari2012mapping} and offers valuable insights for enhancing human-computer interaction~\cite{kramer2012human}.  

While facial expression research is extensive~\cite{keshari2019emotion,seikavandi2023gaze,zhang2023,seikavandi2024modeling}, the temporal dynamics of mimicry remain underexplored. Studies have primarily focused on static aspects, overlooking how expressions align and evolve over time—a key factor in emotional contagion, where individuals unconsciously synchronize expressions.  

Personality traits like Extraversion and Agreeableness influence mimicry responses~\cite{hess2013}, yet their impact on temporal dynamics needs further study.  

To bridge these gaps, this study examines the temporal dynamics of facial mimicry during emotion processing using Action Units (AUs)~\cite{ekman2002}. Dynamic Time Warping (DTW) enables us to analyze temporal alignment, revealing nuanced patterns missed in static analyses~\cite{raducanu2014}.  

These findings can inform affective computing systems—such as virtual agents or telepresence applications—by leveraging personality-linked mimicry patterns to mirror users’ expressions in real time. This approach enhances emotional resonance and engagement, particularly in remote interactions without full physical presence.  

Our research questions are:  

\begin{enumerate}  
    \item How do personality traits influence the temporal dynamics of facial mimicry and emotion recognition?  
    \item What is the relationship between the temporal alignment of facial mimicry and the congruence between perceived and felt emotions?  
    \item How do different emotions affect the temporal patterns of facial mimicry and recognition accuracy?  
\end{enumerate}  

By addressing these questions, we aim to advance the understanding of facial mimicry in emotion processing, with implications for social cognition, clinical interventions, and technology design.


\section{Background}  

Facial mimicry plays a vital role in human communication as a nonverbal channel for emotional exchange. It involves the automatic imitation of others' facial expressions, enhancing emotion recognition and empathy~\cite{oberman2007face,ponari2012mapping,balconi2016empathy}. Studying facial mimicry provides insights into social cognition, affective processes, and neural mechanisms.  

\subsection{Facial Mimicry and Emotion Recognition}  

The \textit{matched motor hypothesis} suggests that observing facial expressions activates corresponding motor representations, aiding emotion recognition through embodied simulation~\cite{niedenthal2006}. This supports the \textit{action-perception loop}, where perceiving expressions trigger motor responses that refine emotional understanding~\cite{chartrand1999}. Mimicry deficits are linked to impaired recognition, especially in negative moods~\cite{hatfield1994,chartrand1999}. The \textit{facial feedback hypothesis} proposes that facial actions amplify emotional experiences, forming a feedback loop between expression and emotion~\cite{wheatley2022,keltner1999}. Mimicry is largely automatic, integrating cues like voice and movement for complete emotional representation~\cite{balconi2007}.  

\subsection{Neural Mechanisms of Facial Mimicry}  

Facial mimicry involves the \textit{mirror neuron system}, which activates during both observation and execution of expressions, facilitating emotional understanding~\cite{iacoboni2005}. Brain imaging links emotional expression observation to empathy and emotion processing areas~\cite{schulte2008}, though the underlying neural pathways remain partially understood.  

\subsection{Personality Traits and Emotion Processing}  

Personality traits from the Big Five Inventory—Openness, Conscientiousness, Extraversion, Agreeableness, and Neuroticism—affect emotion recognition and mimicry. Extraverts, attuned to social cues, exhibit greater mimicry~\cite{personality2021,mccrae2008}, while Neuroticism is associated with a bias toward negative emotions, impairing accuracy~\cite{mimicry_empathy2019}.  

\subsection{Action Units in Emotion Recognition}  

Facial mimicry influences emotional communication, with AUs from the Facial Action Coding System (FACS)~\cite{ekman2002} indicating facial expressions~\cite{westermann2024measuring}. AUs, which categorize muscle movements linked to emotions, are detectable via electromyography (EMG) or image-based techniques. They capture subtle changes, revealing micro-expressions that expose concealed emotions~\cite{microexpressions2021}.Recent advances in AU detection, like Zhi et al.~\cite{zhi2020survey}, highlight deep learning models. Automatic AU imitation enhances empathy and recognition~\cite{holland2021facial}. Research by Pfister et al.~\cite{pfister2011} and Song et al.~\cite{song2022} underscores micro-expression recognition's importance, with technological advances improving AU classification~\cite{wang2020,zhang2021,ActionUnitClassification}.  

\subsection{Dynamic Time Warping and Facial Expressions}  

DTW aligns time-dependent sequences, such as facial expressions and AUs. Raducanu and Dornaika~\cite{raducanu2008} showed that dynamic recognition of facial expressions using DTW outperforms static image analysis. Khan et al.~\cite{khan2019} applied DTW to measure AU similarities, enhancing emotion classification. Zhao et al.~\cite{zhao2018} optimized K-Nearest Neighbor classification with DTW, improving facial expression analysis. Yang et al.~\cite{yang2020} demonstrated DTW's efficacy in emotion recognition, boosting classification accuracy.  

\subsection{Research Gap}  

Despite extensive research on static mimicry, temporal dynamics remain underexplored. AUs provide a granular approach to mimicry analysis. This study bridges gaps by using AUs and DTW to examine the temporal alignment of expressions. We analyze AU divergence and its correlation with personality traits and discrepancies between felt and perceived emotions, offering deeper insights into individual differences in emotion recognition.

\section{Methods}  

\subsection{Participants}  
We recruited 73 participants (52 males, 21 females; mean age $27.4 \pm 6$ years) with normal or corrected-to-normal vision and neurotypicality ( diagnosed with neurodivergent conditions such as autism, ADHD, or dyslexia). Informed consent was obtained per university ethical guidelines.  

\subsection{Experimental Design and Procedure}  
Participants completed 88 trials (4 practice, 84 main) using videos from the CREMA-D dataset~\cite{cao2014crema}, which features actors expressing six emotions at varying intensities.  

The CREMA-D dataset provides demographic diversity with 91 actors (48 male, 43 female) aged 20–74, covering six basic emotions (Anger, Disgust, Fear, Happy, Neutral, Sad) at four intensity levels. We selected 84 clips representing diverse combinations to ensure a balanced emotional and demographic context, supporting generalizability.

To simulate face-to-face dialogue, a written scenario was provided before each video to enhance engagement and emotional alignment. Participants imagined conversing with the individuals in the videos, and responding to their emotions.  

Facial expressions were recorded at up to 40 fps using a webcam, synchronized with stimuli via Lab Streaming Layer (LSL is a framework for real-time data synchronization and integration across devices). AUs were extracted using OpenFace 2.0~\cite{baltruvsaitis2016openface}. Before trials, participants completed the Big Five Inventory (BFI-44)~\cite{john1999} to assess personality traits. After each video, they rated their \emph{perceived} and \emph{felt} emotions on 9-point Likert scales for valence and arousal.  

The 9-point scale was chosen for its higher resolution and sensitivity in capturing subtle emotional variations, enhancing reliability in analyzing complex affective states~\cite{lang2019affective,benitez2022likert}.  

\begin{figure}[htb]  
    \centering  
    \includegraphics[width=0.45\textwidth]{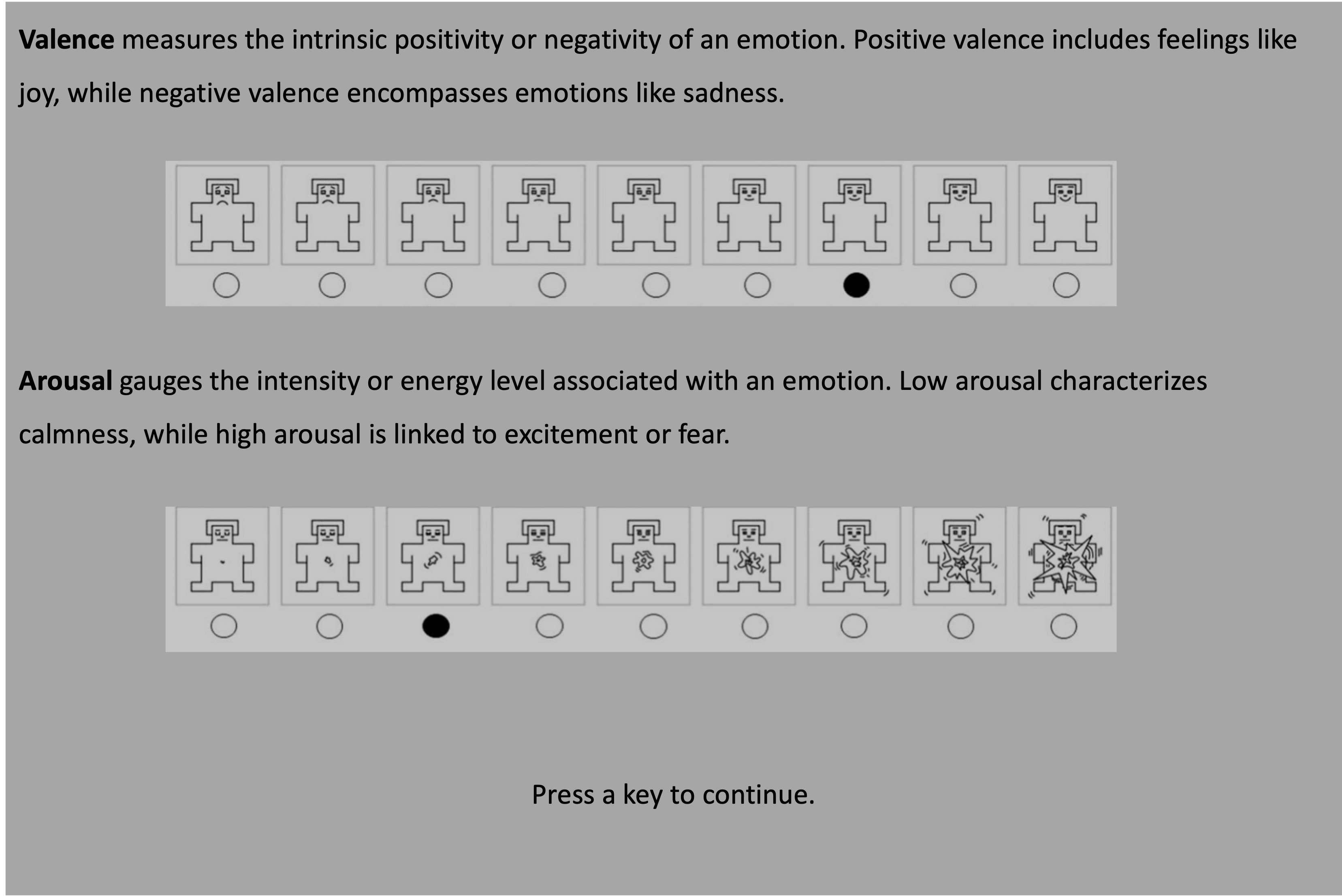}  
    \caption{The 9-point Likert scales for valence and arousal ratings.}  
    \label{fig:likert_scales}  
\end{figure}

The scales are shown in Figure~\ref{fig:likert_scales}.  

\subsection{Data Preprocessing and Feature Extraction}  
Facial Action Units (AUs) were extracted from both participant and stimulus videos using OpenFace 2.0~\cite{baltruvsaitis2016openface}. This process captured 17 key AUs critical for analyzing facial mimicry and emotional expressions. These AUs, listed in Table~\ref{tab:extracted_aus}, provide a comprehensive representation of facial muscle movements, enabling a detailed and nuanced evaluation of mimicry dynamics.

\begin{table}[htb]
\centering
\caption{Action Units (AUs) and Descriptions}
\label{tab:extracted_aus}
\begin{tabular}{|c|l|c|l|}
\hline
\textbf{AU} & \textbf{Description}        & \textbf{AU} & \textbf{Description}      \\ \hline
AU01        & Inner Brow Raiser          & AU14        & Dimpler                  \\ \hline
AU02        & Outer Brow Raiser          & AU15        & Lip Corner Depressor     \\ \hline
AU04        & Brow Lowerer               & AU17        & Chin Raiser             \\ \hline
AU05        & Upper Lid Raiser           & AU20        & Lip Stretcher           \\ \hline
AU06        & Cheek Raiser               & AU23        & Lip Tightener           \\ \hline
AU07        & Lid Tightener              & AU25        & Lips Part               \\ \hline
AU09        & Nose Wrinkler              & AU26        & Jaw Drop                \\ \hline
AU10        & Upper Lip Raiser           & AU45        & Blink                   \\ \hline
AU12        & Lip Corner Puller          &             &                          \\ \hline
\end{tabular}
\end{table}

DTW~\cite{muller2007dynamic} was applied to align AU intensity sequences, quantifying temporal mimicry between participants and stimuli.  

\begin{figure}[htb]  
    \centering  
    \includegraphics[width=0.45\textwidth]{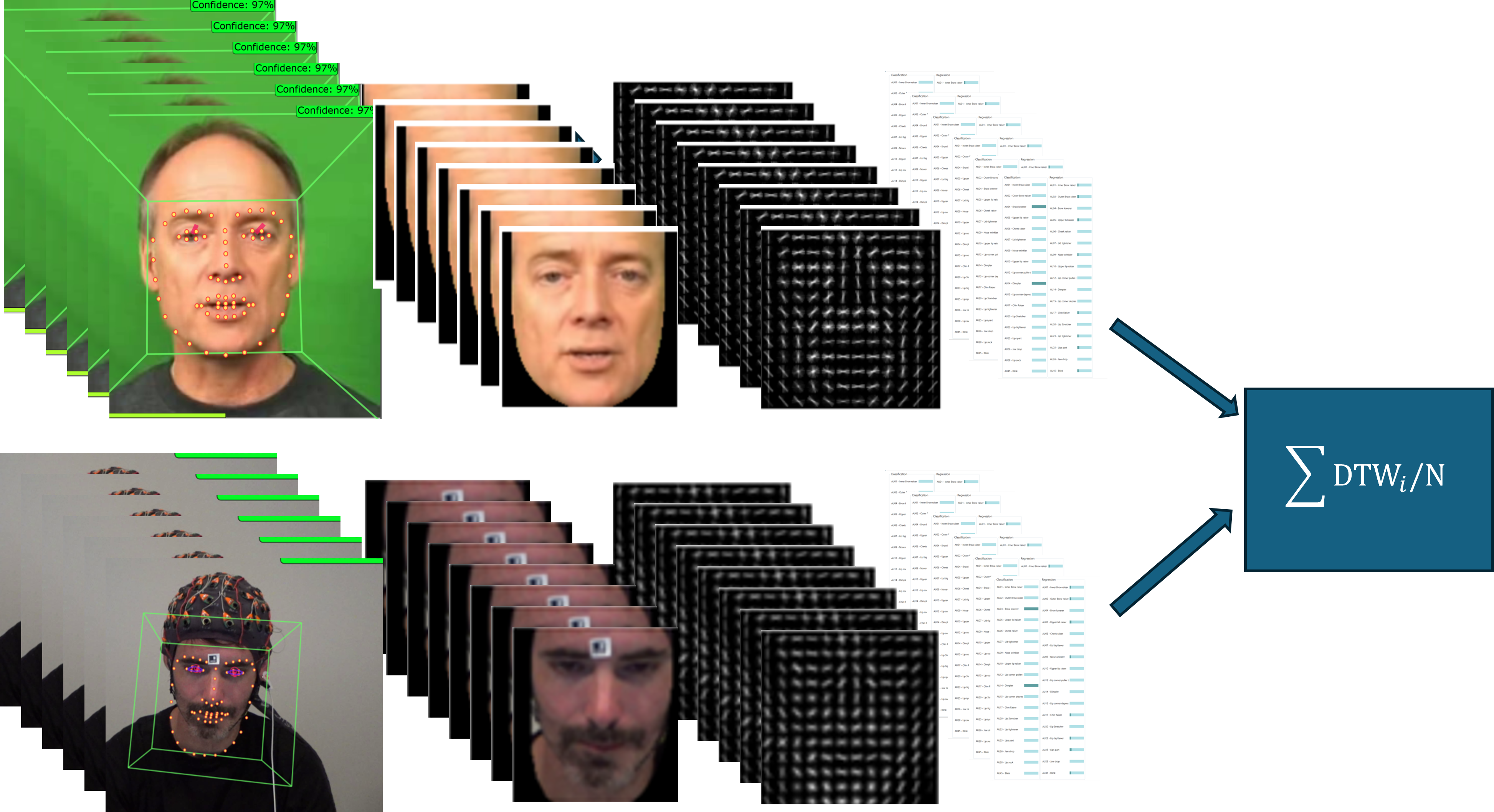}  
    \caption{Framework for similarity detection using DTW on AUs. DTW aligns sequences between participants and stimuli, accounting for temporal variations.}  
    \label{fig:similarity}  
\end{figure}  

DTW handles potential delays in mimicry reactions, capturing subtle dynamic patterns missed by static methods. DTW distance, representing dissimilarity between participant and stimulus AUs, is calculated as:  

\begin{equation}  
\text{DTW}(A, B) = \min_{\phi} \sum_{i=1}^{n} |A(i) - B(\phi(i))|,  
\end{equation}  

where $A$ and $B$ are AU intensity sequences, $n$ is the length of $A$, $\phi$ is the warping function mapping indices of $A$ to $B$, and $|A(i) - B(\phi(i))|$ is the absolute difference between the $i$-th element of $A$ and the $\phi(i)$-th element of $B$.  

Mean and standard deviation of DTW distances were calculated for each trial to capture individual variability.  

The Davies-Bouldin Index (DBI)~\cite{davies1979} was used to evaluate the clustering quality of perceived emotions in the valence-arousal space. Lower DBI values indicate better clustering, assessing the alignment of perceived and felt emotions.  

\section{Statistical Analysis}  

\subsection{Variables and Measures}  
Key variables analyzed:  
\begin{itemize}  
    \item \textbf{Sympathy}: Euclidean distance between perceived and felt emotions in the valence-arousal space, indicating alignment.  
    \item \textbf{Facial Mimicry Divergence}: Average DTW scores across AUs per trial, capturing temporal alignment between participant and stimulus expressions.  
    \item \textbf{Personality Traits}: Scores on Big Five dimensions—Openness, Conscientiousness, Extraversion, Agreeableness, Neuroticism~\cite{john1999}.  
    \item \textbf{Emotion Recognition Performance}: DBI for each emotion category, where lower values indicate better clustering of perceived emotions.  
\end{itemize}

\subsection{Statistical Analyses}

Pearson's correlation and multiple linear regression models were employed to analyze relationships between variables. Statistical significance was set at $\alpha = 0.05$, and Bonferroni corrections were applied for multiple comparisons.

\subsubsection{Correlation Analysis}

Pearson's correlations ($r$) were calculated between:

\begin{itemize}
    \item Sympathy and personality traits (E, C, N, O, A).
    \item Sympathy and facial mimicry divergence (DTW scores).
    \item Emotion recognition performance, DBI and facial mimicry divergence.
    \item Relationships between personality traits and facial mimicry divergence.
\end{itemize}

\vspace{-5pt}
\subsection{Results}

\subsubsection{Sympathy, Personality, and facial mimicry divergence}
Table \ref{tab:correlations} shows which variables correlate significantly with sympathy. Pearson's correlation revealed significant positive correlations between sympathy and perceived emotion arousal ($r = 0.24$, $p < 0.001$), as well as facial mimicry divergence ($r = 0.12$, $p = 0.001$). Sympathy was negatively correlated with perceived emotion valence ($r = -0.17$, $p < 0.001$), indicating higher sympathy for more negative emotions. Positive correlations with personality traits were observed for Agreeableness ($r = 0.10$, $p = 0.005$) and Extraversion ($r = 0.08$, $p = 0.032$).

\begin{table}[h]
\centering
\caption{Significant Correlation Coefficients for Sympathy}
\begin{tabular}{lcc}
\toprule
\textbf{Variable} & \textbf{$r$} & \textbf{$p$-value} \\
\midrule
Perceived Emotion Valence      & $-0.17$ & $< 0.001$ \\
Perceived Emotion Arousal      & $0.24$  & $< 0.001$ \\
Extraversion (E)               & $0.08$  & $0.032$   \\
Agreeableness (A)              & $0.10$  & $0.005$   \\
Facial Mimicry Divergence      & $0.12$  & $0.001$   \\
\bottomrule
\end{tabular}
\label{tab:correlations}
\end{table}

\subsubsection{Emotion Recognition Performance and facial mimicry  divergence}

Correlation analyses (Table \ref{tab:correlations_dbi}) showed that DBI scores were negatively correlated with perceived emotion valence ($r = -0.30$, $p < 0.001$) and felt emotion valence ($r = -0.25$, $p < 0.001$), indicating better performance with more positive emotions. No significant correlation was found between DBI scores and facial mimicry  ($r = 0.05$, $p = 0.12$).

\begin{figure}[htb]
    \centering
    \includegraphics[width=0.48\textwidth]{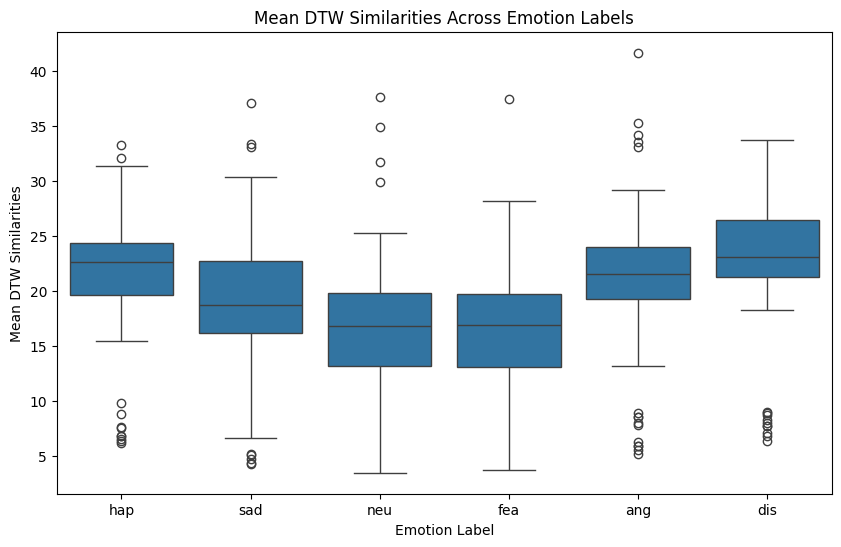}
    \caption{Mean DTW Divergence Across Emotion Labels. The boxplot shows DTW divergence distributions with lower values indicating greater temporal alignment.}
    \label{fig:dtw_similarities}
\end{figure}

\begin{table}[h]
\centering
\caption{Correlation Coefficients between DBI Scores and Other Variables}
\begin{tabular}{lcc}
\toprule
\textbf{Variable} & \textbf{$r$} & \textbf{$p$-value} \\
\midrule
Perceived Emotion Valence      & $-0.30$ & $< 0.001$ \\
Perceived Emotion Arousal      & $0.14$  & $0.004$   \\
Felt Emotion Valence           & $-0.25$ & $< 0.001$ \\
Facial Mimicry divergence      & $0.05$  & $0.12$    \\
\bottomrule
\end{tabular}
\label{tab:correlations_dbi}
\end{table}

\subsection{ANOVA and Post-hoc Tests}

An ANOVA revealed significant effects of emotion labels on both DTW divergences ($F(5,357) = 7.12$, $p < 0.001$) and DBI scores ($F(5,357) = 12.36$, $p < 0.001$). Post-hoc Tukey HSD tests (Table~\ref{tab:tukey_dtw} and Table~\ref{tab:tukey_dbi}) showed that 'Happy' emotions elicited higher DTW divergence compared to 'Fear' and lower divergence for 'Fear' compared to 'Anger' (Figure~\ref{fig:dtw_similarities}). Similarly, DBI scores indicated better emotion recognition for 'Happy' compared to negative emotions like 'Anger' and 'Disgust'. These findings highlight that positive emotions, such as 'Happy,' lead to stronger mimicry and better recognition performance compared to negative emotions.

\begin{table}[h]
\centering
\caption{Tukey HSD Post-hoc Test for DTW Divergence across Emotion Labels}
\begin{tabular}{lccc}
\toprule
\textbf{Comparison} & \textbf{Mean Diff.} & \textbf{$p$-value} & \textbf{Significance} \\
\midrule
Anger vs. Fear   & $-4.40$ & $0.009$ & ** \\
Anger vs. Neutral& $-4.14$ & $0.016$ & *  \\
Fear vs. Happy   & $4.71$  & $0.004$ & ** \\
Disgust vs. Fear & $-5.68$ & $< 0.001$ & *** \\
\bottomrule
\multicolumn{4}{l}{Significance levels: * $p < 0.05$, ** $p < 0.01$, *** $p < 0.001$}
\end{tabular}
\label{tab:tukey_dtw}
\end{table}

\begin{table}[h]
\centering
\caption{Tukey HSD Post-hoc Test for DBI Scores across Emotion Labels}
\begin{tabular}{lccc}
\toprule
\textbf{Comparison} & \textbf{Mean Diff.} & \textbf{$p$-value} & \textbf{Significance} \\
\midrule
Anger vs. Happy   & $-4.04$ & $0.0004$ & *** \\
Disgust vs. Happy & $-4.98$ & $< 0.001$ & *** \\
Fear vs. Happy    & $-6.08$ & $< 0.001$ & *** \\
Fear vs. Sad      & $-3.57$ & $0.003$ & ** \\
\bottomrule
\multicolumn{4}{l}{Significance levels: ** $p < 0.01$, *** $p < 0.001$}
\end{tabular}
\label{tab:tukey_dbi}
\end{table}

\section{Discussion}  

This study provides insights into the dynamic aspects of facial mimicry and its links to personality traits and emotional alignment. Using DTW to analyze AU sequences, we identified temporal alignment patterns that were not evident in static analyses.  

The positive correlation between facial mimicry divergence and sympathy support the facial feedback hypothesis~\cite{wheatley2022}, indicating closer mimicry enhances emotional alignment. This aligns with prior findings that mimicry boosts empathy and emotional understanding~\cite{hatfield1994,dimberg2000}.  

Personality traits like Extraversion and Agreeableness positively correlated with sympathy and mimicry divergence. Extraverts’ stronger mimicry and emotion recognition likely stem from their social responsiveness~\cite{mccrae2008,personality2021}. Similarly, Agreeableness, associated with empathy and prosocial behavior~\cite{graziano2007agreeableness}, showed positive links with mimicry and sympathy.  

No significant relationship was found between Neuroticism and mimicry or emotion recognition, contrasting studies suggesting heightened sensitivity to negative stimuli~\cite{mimicry_empathy2019}, possibly due to limited Neuroticism variability or contextual factors.  

Emotion recognition was more accurate for positive emotions (e.g., happiness) than negative ones (e.g., fear, anger), supporting findings that positive emotions are easier to identify~\cite{wang2023increasing}. The lack of a significant link between mimicry divergence and DBI scores suggests that mimicry enhances emotional alignment but may not universally improve emotion recognition across categories.  

By integrating personality traits and temporal facial expression analysis, our study underscores the importance of temporal dynamics in emotion research~\cite{scherer2009emotions}.  

These findings highlight mimicry's role in social cognition and empathy, with potential applications for improving emotion recognition in populations with deficits (e.g., autism spectrum disorders)~\cite{bal2010emotion}. Considering personality differences could enhance social interaction models and inform the design of affective computing and social robotics systems for natural interactions~\cite{david2022acceptability}.  

Limitations include a homogeneous sample, potential biases in self-reported emotions, and AU measure validity. The gender imbalance in our sample may affect generalizability, as gender influences empathic attitudes and mimicry~\cite{hess2013}. Future studies should include more balanced samples to explore gender differences in emotional processing. Further research should investigate diverse populations, incorporate physiological measures like EMG, and explore faster AU detection methods.


\section{Conclusion}

We demonstrated that greater facial mimicry similarity is associated with higher sympathy, supporting the facial feedback hypothesis. Extraversion and Agreeableness positively correlated with sympathy and mimicry, highlighting individual differences. 


Despite modest variance explained, these findings advance understanding of mimicry's temporal dynamics in emotion processing. 

By integrating dynamic analysis like DTW with detailed facial measures, we uncover intricate mechanisms underlying emotion recognition and social interaction. insights can guide the development of more empathetic and adaptive virtual agents, which could simulate facial mimicry to enhance human-agent interactions.
.

\section{Ethical Impact Statement}

This research investigates the temporal dynamics of facial mimicry in emotion processing by analyzing AUs, which capture subtle facial muscle movements related to emotions. While AU analysis provides valuable insights into emotional alignment and social interaction, it also raises ethical considerations regarding privacy and the autonomy of participants.

Informed consent was obtained from all participants, who were briefed on the study's objectives, data usage, and their right to withdraw at any time without consequences. To protect participant privacy, we extracted AUs from face videos and only used anonymized AU data for analysis. This approach preserves essential aspects of emotional expression while significantly reducing the sensitivity and identifiability of the data compared to raw video footage.

AUs capture fine-grained facial movements, including subtle and often unconscious expressions known as micro-expressions, which can involuntarily convey an individual’s emotional state. Although these micro-expressions were not the primary focus of our analysis, their potential to reveal private emotional information underscores a need for caution in interpreting such data, particularly in applications where individuals cannot consciously control the emotional cues they display. Using mimicry for behavioral predictions, such as assessing emotional alignment or social intentions, raises concerns about potential overreach, misinterpretation, or misuse in sensitive contexts like surveillance or biased decision-making. Like any technology, facial mimicry analysis can drive positive innovation when used responsibly or pose risks if misused. We emphasize the importance of robust ethical guidelines to ensure transparent, fair, and responsible applications that prioritize individual rights and societal well-being.

The findings from this study have potential applications in enhancing human-computer interaction, social robotics, and clinical interventions. However, given the automatic and largely unconscious nature of many facial expressions, there is a risk of misuse in scenarios that might limit an individual's ability to manage the emotional signals they disclose. Additionally, automated frameworks like OpenFace, while powerful, may have inherent biases that could affect AU predictions across diverse demographic groups. Acknowledging these biases is crucial to ensure fairness and reliability in emotion recognition applications. Careful, ethically aligned deployment of these findings is therefore essential to uphold participants' rights to privacy and autonomy.

We have prioritized ethical data handling and anonymization protocols throughout this study, ensuring that the analysis of facial mimicry is conducted with respect for individual privacy and ethical standards. This commitment aligns with our objective to advance affective computing while maintaining a balanced approach to innovation and privacy protection.

\bibliographystyle{ieee}
\bibliography{egbib}
\end{document}